\documentstyle [aps,multicol,prb]{revtex}
\begin{document}                                                    
\draft                                                      
\title {Mixed charge-spin response functions of an arbitrarily  
 polarized electron gas}
\author{Sudhakar Yarlagadda}
\address{Saha Institute of Nuclear Physics, 1/AF Bidhannagar, Calcutta}
\date{\today}
\maketitle
\begin{abstract}
In this paper, using different approaches we demonstrate the equality 
of the two mixed charge-spin 
response functions of a spin-polarized electron gas 
when orbital effects are negligible.
  Within a generalized STLS approximation
we show that the two mixed responses are equal.
We also present arguments for the equality
of the two dynamic responses by considering a symmetry
of the effective 
screened interaction between
two opposite spin electrons.
 Furthermore, using the reflection symmetry of the system
 and the fact that the hamiltonian
is real we prove
rigorously that the two responses coincide identically.
\end{abstract}

                                                            
\begin{multicols}{2}
\section{INTRODUCTION}                                
An electron gas (EG) can be macroscopically 
characterized in terms of its
response functions. For an unpolarized EG in a uniform
positive background, using many-body local fields
to account for the vertex corrections associated with
the charge and spin density fluctuations,
it has been shown
in Refs.\ \onlinecite {KO,VS,YG1}
that only the charge response
 and the longitudinal spin response functions are
sufficient to obtain the self-energy and the effective interaction
between electrons. Again using
the many-body local field formalism, for a
system  that is polarized it was 
however found that one has
to also have knowledge of 
the two mixed charge-spin response functions
(i.e., the charge response to
an external magnetic field and the spin response to an external
electric potential)
 and the transverse spin response \cite {YG2}.
Earlier on Kim et al. \cite{kim} treated the
 two charge-spin mixed response
 functions 
 within a random phase approximation and found
them to be identical.
More recently, Yi and Quinn \cite {YQ}
studied these two responses 
in an uniform EG which is polarized by a  DC magnetic
field but which has negligible orbital effects.
These authors conjectured that
the two mixed response
 functions, in general, can be different.

In this paper we demonstrate
the equality of the two mixed response functions
in the polarized system considered in Ref. \onlinecite{YQ}.
In Sec. II, we use arguments of increasing sophistication/rigor
to establish the coincidence of the two responses.
 We first present arguments for
the equality of the responses within a
 generalized STLS approximation
\cite{NgSingwi}. Next we show that
 the mixed charge-spin responses are equal by 
exploiting a symmetry of the effective screened interaction between
two opposite spin electrons. 
 Lastly by invoking the reflection symmetry of the system and the fact
that the hamiltonian is not only hermitian but also real,
 we prove that
the two responses are exactly identical.
 Finally in Sec. III we
discuss the behavior of the mixed responses in the limit of
large momentum or large frequency and also present our
conclusions.

\section{SYMMETRY BASED EQUALITY OF RESPONSES}        

In the analysis that follows we will ignore the effect
of the induced magnetization.
We will first
derive expressions for the
 charge, the spin, and the two
mixed response functions. 
Using arguments similar to those used  in 
Ref.\ \onlinecite {YG2}
we note that the potential felt by an electron of spin
$\sigma$ when an external 
potential is applied is given by 
\begin{eqnarray}                                          
\Phi ^{\sigma}(\vec{q}, \omega )  =
\Delta n ^{\sigma} / \chi^{\sigma}_{0}=
\phi ^{\sigma} +
&& v(q)  \left [
 \Delta n^{-\sigma}(1 - 2 G^{\sigma , -\sigma} ) 
 \right .
\nonumber \\
  && 
 \left . +
 \Delta n^{\sigma}(1 - 2 G^{\sigma  , \sigma} )
\right ] ,
\label{phiCS}
\end{eqnarray}                                          
where
$v(q)$ is the Coulombic interaction between two electrons.
  Furthermore 
$ \chi^{\sigma}_{0}$ is the non-interacting polarizability,
 $\Delta n^{\sigma}$
is the density fluctuation, and 
 $G^{\sigma , \sigma ^{\prime}}$ are the many-body local fields
and are all functions of $\vec{q}$ and $\omega$.
In Eq.\ (\ref{phiCS}) for an applied spin
 symmetric [spin antisymmetric]
electric potential $V_{ext} (\vec{q}, \omega )$ 
[magnetic field $H_{ext} (\vec{q}, \omega )$  ],
$ \phi ^{\sigma}
(\vec{q}, \omega )  
 = eV _{ext}$ $ [g \sigma \mu _{B} H _{ext}]$.
On defining $G_{\pm} ^{\sigma} \equiv
 G^{\sigma , \sigma}
 \pm G^{\sigma ,-\sigma}$
we obtain, in agreement with
 Ref.\ \onlinecite{YQ}
 the following expressions for the
charge, the longitudinal spin, and the two mixed
 charge-spin  response functions: 
\begin{eqnarray}                                          
 \chi_{C}(\vec{q} ,\omega ) && = 
\frac{\Delta n^{\uparrow} 
+\Delta n^{\downarrow}}{eV_{ext}} 
\nonumber \\
&& =\frac{(\chi ^{\uparrow}_{0} +\chi^{\downarrow}_{0} ) +
2 v(q) \chi ^{\uparrow}_{0} \chi^{\downarrow}_{0} (
 G^{\downarrow}_{-}
+ G^{\uparrow}_{-} )}{\cal{D}} ,
\label{chiC}
\end{eqnarray}                                          
\begin{eqnarray}                                          
\frac{ \chi_{S}(\vec{q} ,\omega )}{-\mu_{B}^2} && =
\frac{\Delta n^{\uparrow} 
-\Delta n^{\downarrow}}{\mu_{B} H_{ext}}
\nonumber \\
&& =\frac{(\chi ^{\uparrow}_{0} +\chi^{\downarrow}_{0} ) +
2 v(q) \chi ^{\uparrow}_{0} \chi^{\downarrow}_{0} (
 G^{\downarrow}_{+}
+ G^{\uparrow}_{+} -2 )}{\cal{D}} ,
\label{chiS}
\end{eqnarray}                                          
\begin{eqnarray}                                          
 \chi_{CS}^{V}(\vec{q} ,\omega ) && =
\frac{\Delta n^{\uparrow} 
-\Delta n^{\downarrow}}{eV_{ext}} 
\nonumber \\
&& =\frac{(\chi ^{\uparrow}_{0} -\chi^{\downarrow}_{0} ) +
2 v(q) \chi ^{\uparrow}_{0} \chi^{\downarrow}_{0} (
 G^{\downarrow}_{+}
- G^{\uparrow}_{+} )}{\cal{D}} ,
\label{chiCSV}
\end{eqnarray}                                          
and
\begin{eqnarray}                                          
 \chi_{CS}^{H}(\vec{q} ,\omega ) && =
\frac{\Delta n^{\uparrow} 
+\Delta n^{\downarrow}}{\mu_{B} H_{ext}}
\nonumber \\
&& =\frac{(\chi ^{\uparrow}_{0} -\chi^{\downarrow}_{0} ) +
2 v(q) \chi ^{\uparrow}_{0} \chi^{\downarrow}_{0} (
 G^{\downarrow}_{-}
- G^{\uparrow}_{-} )}{\cal{D}} ,
\label{chiCSH}
\end{eqnarray}                                          
where
\begin{eqnarray}                                          
{\cal{D}} = && 
1-
 v(q) \chi ^{\uparrow}_{0} [ 1 -
 G^{\uparrow}_{+} -
 G^{\uparrow}_{-} ]
 - v(q) \chi ^{\downarrow}_{0} [ 1 -
 G^{\downarrow}_{+} -
 G^{\downarrow}_{-} ]
\nonumber \\
 && 
- 2 v(q) ^2 \chi ^{\uparrow}_{0} \chi^{\downarrow}_{0} [
 G^{\uparrow}_{-} (1 -
 G^{\downarrow}_{+})
 + G^{\downarrow}_{-} (1 -
 G^{\uparrow}_{+})].
\label{calD}
\end{eqnarray}                                       
Thus from Eqs.\ (\ref{chiCSV}) and (\ref{chiCSH})
we see that 
the mixed response functions
are equal to each other  if and only if
 $G^{\uparrow , \downarrow} (q, \omega)  =
 G^{\downarrow , \uparrow} (q, \omega )$.

We will now argue, within the generalized STLS approximation
formulated by Ng and Singwi \cite{NgSingwi}, that the two mixed
response functions are equal.
Within the generalized STLS scheme the density fluctuations
$\Delta n ^{\sigma} (\vec{q} , \omega )$ induced in the EG
 by an external spin dependent potential 
$\phi ^{\sigma} (\vec{q}, \omega )$ are given by
 (see Ref.\ \onlinecite{NgSingwi})
\begin{eqnarray}                                          
\Delta n ^{\sigma} / \chi^{\sigma}_{0} (\vec{q} , \omega ) =
\phi ^{\sigma} +
 v(q) && \left [
  \Delta n^{-\sigma} \left \{ 1 - 2 {\cal G}^{\sigma , -\sigma} (q) 
 \right \}
  \right . 
\nonumber \\ && +  \left . \left \{
 \Delta n^{\sigma}(1 - 2 {\cal G}^{\sigma  , \sigma} (q) \right \}
 \right ] ,
\label{STLS}
\end{eqnarray}                                          
where 
${\cal G}^{\sigma, \sigma^{\prime}}(q)$ are the
 {\it static} many-body local fields. 
Here we
would like to reiterate the known fact that the static
nature of
the local fields
${\cal G}^{\sigma, \sigma^{\prime}}(q)$  in the 
STLS scheme is a consequence
of the approximation made in obtaining the time-dependent
 two-particle
distribution function and is not a matter of definition
as might appear from Eq. (\ref{STLS}).
 From the above considerations
[using Eqs. (\ref{phiCS}) and (\ref{STLS})]
 we see that the mixed responses
within the generalized STLS approach are obtained by replacing
$ G^{\sigma, \sigma^{\prime}}(q , \omega )$ with
${\cal G}^{\sigma, \sigma^{\prime}}(q)$ 
in the expressions for the mixed responses given by Eqs. (\ref{chiCSV})
and (\ref{chiCSH}). 
It has been shown, by Ng and Singwi 
 that 
$ V^{\sigma \sigma^{\prime}}_{\rm{eff}}(q) 
 \equiv  v(q)[1- 2{\cal G}^{\sigma, \sigma^{\prime}}(q)]$
 is related to the pair correlation function in real space
as follows
\begin{equation}                                          
V^{\sigma \sigma^{\prime}}_{\rm{eff}}(r) = 
- \int_{r} ^{\infty} \frac{d v(R)}{d R}
g_{\sigma \sigma^{\prime}}(R) d R.
\label{Veffgreltn}
\end{equation}                                          

We will now demonstrate that
$g_{\sigma  -\sigma } ( \vec{r} ) =
g_{- \sigma  \sigma } ( \vec{r} )$.
  Firstly we observe, for an arbitrarily spin polarized EG in any
 dimension, that due to reflection symmetry
the static pair correlation
 function $g_{\sigma  \sigma ^{\prime}} ( \vec{r} )$
satisfies the relationship 
$g_{\sigma  \sigma ^{\prime}} ( \vec{r} )
= g_{\sigma  \sigma ^{\prime}} ( - \vec{r} )$.
Next we note that 
the pair correlation function 
$g_{\sigma - \sigma } ( \vec{r} )$
is related to the instantaneous density-density
 correlation function through
the relation
\begin{equation}                                          
g_{\sigma  - \sigma } ( \vec{r} ) 
 =
 \frac {
<0|\rho ^{\sigma} (\vec{r} + \vec{r} ^{\prime})
\rho ^{{- \sigma}} ( \vec{r} ^{\prime}) |0> }
{N^{\sigma} N ^{- \sigma } } ,
\label{paircor}
\end{equation}                                          
where $N^{\sigma}$ is the number of spin 
$\sigma$ particles  in a system
of unit volume and
$\rho ^{\sigma} (\vec{r})$
is the density operator in real space. 
Then, since the system has reflection symmetry and translational
invariance,
it follows that 
$g_{\sigma  -\sigma } ( \vec{r} ) =
g_{- \sigma  \sigma } ( \vec{r} )$ \cite{alternate}.
Then from Eq. (\ref{Veffgreltn}) it is clear that
 $ {\cal G}^{\sigma, -\sigma}(q)
 = {\cal G}^{- \sigma, \sigma}(q)$ and thus we see that the two mixed
responses are equal within the generalized STLS scheme.
Here it should be pointed out that the many-body local fields 
${\cal G}^{\sigma, \sigma^{\prime}}(q)$  in the
generalized STLS approximation
 are not functions of frequency and thus are not
the same as the true static many-body local fields
$G^{\sigma, \sigma^{\prime}}(q, 0)$.

 We will now proceed to give arguments for 
the  equality of the two dynamic many-body local fields
 $G^{\uparrow , \downarrow} (q, \omega)$ and
 $ G^{\downarrow , \uparrow} (q, \omega )$.
 To this end we will derive the effective  screened
potential 
$W_{-\sigma \sigma}$
 felt by a spin $-\sigma$ electron in the EG
when a spin $\sigma$ electron sets up perturbations
and then show that based on a symmetry of
$W_{-\sigma \sigma}$
one obtains
 $G^{\uparrow , \downarrow} (q, \omega)  =
 G^{\downarrow , \uparrow} (q, \omega )$.
Let $\rho ^{\sigma} _{e}$ be the number density
of the perturbing
 electron setting up 
 density fluctuations
$\Delta n^{\sigma}$ and
$\Delta n^{-\sigma}$.
Then the potential felt by an electron of spin
$\sigma ^{\prime}$
is given by 
\begin{eqnarray*}                                          
\Phi ^{\star} _{\sigma ^{\prime} \sigma } =
\Delta n ^{\sigma ^{\prime}} / \chi^{\sigma ^{\prime}}_{0}=
\phi _{\sigma ^{\prime} \sigma } 
 +
 v(q) &&
 [
\Delta n^{-\sigma}(1 - 2 G^{\sigma ^{\prime} , -\sigma} )
\nonumber \\
 && +
 \Delta n^{\sigma}(1 - 2 G^{\sigma ^{\prime} , \sigma} )
 ] ,
\end{eqnarray*}                                          
where
\begin{equation}                                          
\phi _{\sigma ^{\prime} \sigma }  =
 v(q) \rho^{\sigma} _{e}(1 - 2 G^{\sigma ^{\prime} , \sigma} ) .
\end{equation}                                          
On simplifying the above one gets
\begin{equation}                                          
\Phi ^{\star} _{- \sigma  \sigma } =
 \frac { \phi _{- \sigma  \sigma } } 
 {\cal{D}} ,
\end{equation}                                          
where $\cal{D}$ is given by Eq.\ (\ref{calD}).
Now, as argued in 
Refs.\ \onlinecite{KO} and \onlinecite{YG4}
 for an unpolarized system,
the actual screened potential
between the two electrons
$ W_{\sigma ^{\prime} \sigma }$
 is given by
\begin{equation}                                          
W _{\sigma ^{\prime} \sigma } 
 \rho^{\sigma} _{e} =
 \Phi ^{\star} _{\sigma ^{\prime} \sigma } +
2 v(q) \rho^{\sigma} _{e}
{  G^{\sigma ^{\prime} , \sigma} } .
\label{Phisigmasigma}
\end{equation}                                          
The effective screened potential 
$ W _{\sigma ^{\prime} \sigma } $
obtained in such a fashion also
agrees, for an
infinitesimally polarized system,
 with the results of 
Ref.\ \onlinecite{YG2}.
 Furthermore, in our treatment if
 $ \rho^{-\sigma}_{e}$ is a spectator electron
which experiences an effective screened potential
$W_{-\sigma \sigma} $,
  the interaction energy between the two electrons
$ \rho ^{\sigma} _{e}$ and $ \rho^{-\sigma} _{e}$ is
$W_{-\sigma \sigma} 
\rho^{\sigma} _{e}
\rho^{-\sigma} _{e}$.
 From the fact that the 
 interaction energy between
the two electrons
$ \rho ^{\sigma} _{e}$
and $ \rho^{-\sigma} _{e}$
 is the same
irrespective of whether
$ \rho ^{\sigma} _{e}$ is the perturbing electron
with  $ \rho^{-\sigma} _{e}$ being
 the spectator electron or vice versa,
we have
$W_{\uparrow \downarrow} 
 = W_{\downarrow \uparrow} $.
It then follows from Eq. (\ref{Phisigmasigma}) that
 $G^{\uparrow , \downarrow} (q, \omega)  =
 G^{\downarrow , \uparrow} (q, \omega )$
and hence that
 $\chi_{CS}^{V}(\vec{q} , \omega ) =
 \chi_{CS}^{H}(\vec{q} ,\omega )$.

We also note that the effective screened interaction between two
electrons can be expressed as
\begin{eqnarray}                                          
W_{\sigma \sigma}
 =  && v(q)^2 
\left \{
 ( 1 - G^{\sigma}_{+})^{2} \chi_{C}
 +(G^{\sigma}_{-})^{2} \chi_{S}/(-\mu_{B}^2 )
 \right .
\nonumber \\
 && 
 \left .
 - \sigma ( 1 - G^{\sigma}_{+}) G^{\sigma}_{-} 
[ \chi_{CS}^{V}+ \chi_{CS}^{H}] \right \}
\nonumber \\
+ v(q) && +2 \sigma v(q)^2 \chi ^{-\sigma}_{0}
 ( 1 - G^{\sigma}_{+} + G^{\sigma}_{-} )
\frac {G^{\uparrow \downarrow} -
 G^{\downarrow \uparrow}}
{\cal{D}}  ,
\end{eqnarray}                                          
and
\begin{eqnarray}                                          
W_{\downarrow \uparrow}
 =  v(q)^2 &&
\left \{
 ( 1 - G^{\uparrow}_{+})
 ( 1 - G^{\downarrow}_{+})
 \chi_{C}
 -G^{\uparrow}_{-}
 G^{\downarrow}_{-}
 \chi_{S}/(-\mu_{B}^2 ) \right .
\nonumber \\
 && 
 \left .
 +( 1 - G^{\uparrow}_{+}) G^{\downarrow}_{-} 
\chi_{CS}^{H}
 -( 1 - G^{\downarrow}_{+}) G^{\uparrow}_{-} 
 \chi_{CS}^{V}
\right \}
\nonumber \\
 + v(q) && +2 v(q)^2 \chi ^{\downarrow}_{0}
 ( 1 - G^{\downarrow}_{+} - G^{\downarrow}_{-} )
\frac {G^{\uparrow \downarrow} -
 G^{\downarrow \uparrow}}
{\cal{D}}  .
\end{eqnarray}                                          
When $G^{\uparrow \downarrow} =
 G^{\downarrow \uparrow}$
we see from the above two equations for 
$W_{\sigma \sigma}$ 
and $W_{\downarrow \uparrow}$ 
that the last term vanishes
and the resulting expressions  have  a pleasing symmetry 
in terms of the response functions.

We will now prove that the two responses are identical to
each other by invoking the reflection symmetry and
the real valued property of the hamiltonian.
 For this purpose, we begin with the definitions 
of the mixed charge-spin response functions.
\begin{equation}                                                     
\chi _{CS}^{V}(\vec{q}, t) \equiv  -i \theta (t) 
<0|[S^z_{-\vec{q}} (t),\rho_{\vec{q}}]|0>  ,
\end{equation}                                                       
and
\begin{equation} 
\chi _{CS}^{H}(\vec{q}, t) \equiv 
 - i \theta (t) 
<0| [ \rho_{-\vec{q}} (t), S^z_{\vec{q}} ]|0>   ,
\end{equation}                    
where  $\chi _{CS}^V$
 $(\chi _{CS}^H)$ corresponds to the spin (charge) response
when an external electric potential (magnetic field) is applied.
 Furthermore $\rho_{\vec{q}}$ and $ S^z_{-\vec{q}} $ 
represent respectively the charge and 
spin density fluctuation operators
in momentum space, while $\theta (t)$ stands for the step function.
 From the above definitions one obtains (see Ref.\ \onlinecite{YG2} for
alternate derivations)
  \begin{eqnarray}                                
  \chi _{CS}^{H} ( \vec{q} , \nu )       
   = \sum _{n} && \left \{       
  \frac {                                            
  \langle 0 | \rho _{-\vec{q}}| n \rangle \langle n | S     
  _{\vec{q}}^{z}|                                      
  0 \rangle }                                           
  {\nu -\omega _{n0} + i \eta}
 \right .
 \nonumber \\                                                 
 && ~~~~ \left .
 - \frac {                                     
  \langle 0 | S _{\vec{q}} ^{z} | n \rangle \langle n |    
  \rho _{-\vec{q}}|                                
  0 \rangle }                                    
  {\nu +\omega _{n0} + i \eta}                    
  \right  \}    ,
\label{CSH}
  \end{eqnarray}                                
and
  \begin{eqnarray}                                
  \chi _{CS}^{V} ( \vec{q} , \nu )        
 =  \sum _{n} && \left \{                           
  \frac {                                              
  \langle 0 | S _{-\vec{q}} ^{z}| n \rangle \langle n |     
   \rho _{\vec{q}}|                                   
  0 \rangle }                                             
  {\nu -\omega _{n0} + i \eta}                       
 \right .
 \nonumber \\                                                 
 && ~~~~ \left .
  - \frac {                                                  
\langle 0 | \rho _{\vec{q}}| n \rangle
  \langle n | S _{-\vec{q}} ^{z} | 0 \rangle 
 }         
  {\nu +\omega _{n0} + i \eta}                   
  \right \}  .                                    
\label{CSV}
  \end{eqnarray}                                
                                                 
Next, in the hamiltonian
we note that the kinetic term 
($p^2 _{i} /(2m)$)
and the interaction term
 ($\sum_{i \neq j} v(\vec{r}_{i} -\vec{r}_{j})$) 
have inversion symmetry.
 Furthermore the Zeeman term 
($g \mu _B \vec{\sigma}_{i} \cdot \vec{H} _{ext}$)
also has reflection symmetry because both spin and magnetic field
are pseudo vectors.
 Then, since reflection symmetry holds for all the terms
of the hamiltonian,
we have
 $\chi_{CS}^{V(H)}(\vec{q} ,\omega )
=\chi_{CS}^{V(H)}(-\vec{q} ,\omega )$ and hence we obtain 
 the following equality:
\begin{eqnarray*}                                          
 \chi_{CS}^{V}(\vec{q} , \omega =0 )
= && \chi_{CS}^{H}(\vec{q} ,\omega =0 )
\nonumber \\
= && 
- \sum _{n} 
 \frac {1}{\omega_{n0}} 
  \left \{ 
<0|S^z_{\vec{q}} |n><n|\rho_{-\vec{q}}|0> 
 \right .
\nonumber \\  && ~~~~~~~~~~~~ +
\left . 
<0|\rho_{-\vec{q}}|n> <n|S^z_{\vec{q}} |0> \right \} . 
\end{eqnarray*}                                          
Thus we see that reflection symmetry alone is sufficient to
produce equality
of the two  mixed responses in the static case. 

We will now prove that the two mixed responses coincide in the more
general dynamic case as well. We first note that the hamiltonian is 
real when the external uniform magnetic field is taken to be in 
the z-direction leading to the Zeeman term being real. 
Then the eigenvectors of the hamiltonian
can be taken to be real. Consequently we observe that
$<n|\rho_{\vec{q}}|0> = <0|\rho_{\vec{q}}|n>$ 
and $<0|S^z_{\vec{q}} |n>= <n|S^z_{\vec{q}} |0>$. 
Based on the above equalities and  reflection
symmetry it follows from
 Eqs.\ (\ref{CSH}) and (\ref{CSV}) that
 $\chi_{CS}^{H}(\vec{q} ,\omega )
=\chi_{CS}^{V}(-\vec{q} ,\omega )
=\chi_{CS}^{V}(\vec{q} ,\omega )$.
Here we would like to point out that reflection symmetry holds
even if orbital effects are not negligible. However, when
the orbital effects are included,
the eigenvectors need not be real and hence the mixed responses
need not be equal.

The arguments involving reflection symmetry and the hamiltonian being
real can be generalized to give the equality of the response functions
$  -i \theta (t) <0|[A_{-\vec{q}} (t),B_{\vec{q}}]|0>  $ and
$  -i \theta (t) <0|[B_{-\vec{q}} (t),A_{\vec{q}}]|0>  $ where $A$ and
$B$ are operators. Here, if in one response function
we interchange the time independent operator that couples to the
external field with the time-dependent operator that corresponds to the
response of the system we get the other response function.

\section{DISCUSSION AND CONCLUSIONS}
 
In Ref. \onlinecite{quinn2}, Marinescu and Quinn
 obtained exact expressions
for the many-body local fields of a polarized system in the large
momentum limit or large frequency limit by using approaches
 similar to 
those used
for unpolarized systems
in Refs. \onlinecite{nik} and \onlinecite{zhuawo}.
In the large momentum limit,
 $G^{\uparrow , \downarrow}$ and  
 $ G^{\downarrow , \uparrow}$
 have the same functional dependence on 
the pair correlation functions 
 $g^{\uparrow , \downarrow} (0) $ and
 $ g^{\downarrow , \uparrow} (0) $ respectively and
hence it follows from reflection symmetry that
 $G^{\uparrow , \downarrow} (q \rightarrow \infty , \omega )
 = G^{\downarrow , \uparrow} (q \rightarrow \infty , \omega ) $.
 Furthermore for finite $q$ and $\omega \rightarrow \infty $,
we note that $\chi ^{\sigma} _0 (q , \omega ) 
\sim N^{\sigma} q^2/(m \omega ^2)$ and from Ref. \onlinecite{quinn2}
that
 $G^{\sigma , -\sigma} (q , \omega \rightarrow \infty ) $
is even at least up to order $1/\omega ^2$.
Then, from Eqs. (\ref{chiCSV})--(\ref{calD}) it follows that the
 difference between the two responses is of the form
\begin{eqnarray*}                                          
 \chi_{CS}^{V}(\vec{q} , \omega \rightarrow \infty )
- \chi_{CS}^{H}(\vec{q} ,\omega \rightarrow \infty )
 \sim A/\omega ^4 + B/\omega ^6
+...  ,
\end{eqnarray*}                                         
and is thus an even function of $1/\omega $
 to at least order $1/\omega ^6$.
Next, from Eqs. (\ref{CSH}) and (\ref{CSV}), on using only
reflection symmetry, we note that
\begin{eqnarray*}                                          
  \chi_{CS}^{V} (\vec{q} , \omega \rightarrow \infty )
- && \chi_{CS}^{H}(\vec{q} ,\omega \rightarrow \infty )
\nonumber \\
 &&  
=
2 \sum _{n} 
  \left \{ 
<0|S^z_{\vec{q}} |n><n|\rho_{-\vec{q}}|0>  \right .
\nonumber \\
 &&  ~~~~~~~~
 \left . - <0|\rho_{-\vec{q}}|n> <n|S^z_{\vec{q}} |0> \right \} 
\nonumber \\  && ~~~~~~~~
\times
 \left \{ 
\frac {1}{\omega} +
\frac {\omega_{n0}^2} {\omega ^3} +
\frac {\omega_{n0}^4} {\omega ^5} + ...\right \} .
\end{eqnarray*}                                 
Hence,
 for $\omega \rightarrow \infty $ we see from the above two equations
 that the two mixed responses,
when expanded as a series in powers of $1/\omega $,
coincide at least to order
$1/\omega ^6$.

In conclusion,
within a generalized STLS framework we showed that the
two  charge-spin response functions coincide.
 Furthermore, we also presented 
arguments for their equality in the dynamic
case by considering a symmetry of
the effective screened interaction between two opposite spin electrons. 
Lastly, from the facts that the system has reflection symmetry and that
the hamiltonian is real we established
the equality of the two 
 mixed  response  functions.

Extending the study of mixed responses
 to quantum Hall effect systems
and including orbital effects
could lead to some interesting insights.

\thanks{   The author would like to thank Subrata Ray
and A. Harindranath
 for 
 discussions.}

\end{multicols}

\begin{references}
\bibitem{alternate}
Alternately, by defining the pair distribution function in terms
of the many-body wavefunction, we can also show the equality
$g_{\sigma  \sigma ^{\prime}} ( \vec{r} ) =
g_{\sigma^{\prime}  \sigma } ( \vec{r} )$.
\end{references}
\end{document}